# Loss-difference-induced localization in a non-Hermitian honeycomb photonic lattice

Yuan Feng[1,*], Zhenzhi Liu[1,*], Fu Liu[1,†], Jiawei Yu[1], Shun Liang[1], Feng Li[1], Yanpeng Zhang[1], Min Xiao[2,3] and Zhaoyang Zhang[1,‡]

[1]*Key Laboratory for Physical Electronics and Devices of the Ministry of Education & Shaanxi Key Lab of Information Photonic Technique, School of Electronic Science and Engineering, Faculty of Electronic and Information Engineering, Xi'an Jiaotong University, Xi'an, 710049, China*
[2]*Department of Physics, University of Arkansas, Fayetteville, Arkansas, 72701, USA*
[3]*National Laboratory of Solid State Microstructures and School of Physics, Nanjing University, Nanjing 210093, China*

Non-Hermitian systems with complex-valued energy spectra provide an extraordinary platform for manipulating unconventional dynamics of light. Here, we demonstrate the localization of light in an instantaneously reconfigurable non-Hermitian honeycomb photonic lattice that is established in a coherently-prepared atomic system. One set of the sublattices is optically modulated to introduce the absorptive difference between neighboring lattice sites, where the Dirac points in reciprocal space are extended into dispersionless local flat bands. When these local flat bands are broad enough due to larger loss difference, the incident beam is effectively localized at one set of the lattices with weaker absorption, namely, the commonly seen power exchange between adjacent channels in photonic lattices is effectively prohibited. The current work unlocks a new capability from non-Hermitian two-dimensional photonic lattices and provides an alternative route for engineering tunable local flat bands in photonic structures.

Loss is usually considered to be detrimental for wave manipulations as it leads to energy decay. However, since the introduction of non-Hermitian physics within the context of quantum field theories into optical and other systems, the combination of dispersion and loss in desired manners has inspired considerable new phenomena, particularly, the unusual wave dynamical characteristics arising from the complex-valued eigenvalue spectra [1]-[7], such as non-Hermitian topological properties [8-10], non-Hermitian skin effect [11,12], lasing and optical sensing [13,14], *etc*. Most of the counterintuitive features in non-Hermitian optical settings [15-20] are closely related to the well-known exceptional points [21,22], around which the complex eigenvalues or band structures undergo bifurcation and an abrupt phase transition occurs. To date, the studies of non-Hermitian potentials have been extended to high-dimensional framework, where exceptional rings emerge in momentum space, inside which local flat bands become possible [18,19,23,24]. These pioneering works mainly focused on the exotic wave dynamics around exceptional rings by manipulating the degree of non-Hermiticity.

Recently, local flat bands inside exceptional rings are also theoretically proposed to govern the behaviors of light in distinct manners, for example, chiroptical polarization response [25]. Expanding the research interests from the vicinity of exceptional rings to the ring-surrounded areas will further enrich the capabilities and applications of high-dimensional non-Hermitian systems. Nevertheless, the experimental exploration on the underlying properties of the areas that surrounded by exceptional rings is rarely reported.

One intrinsic property of the dispersionless flat band is to realize field localization. In general, the localization of optical field in periodic photonic structures is achieved by taking advantage of energy flat bands in the whole Brillouin zone, such as the localizations in one dimensional photonic chain [26], photonic Lieb lattices [27,28], and moiré superlattices [29], among others. This is because the non-dispersive flat band guarantees zero group velocity, therefore the wave package will propagate in a dispersionless manner through media. In such a case, however, it is not necessary to have an entire flat band for the field localization, as the Fourier components of the wave package usually cover part of the momentum space, rather than the whole $k$ space.

In this paper, we experimentally demonstrate the field localization in a non-Hermitian honeycomb photonic lattice exhibiting local flat bands (around both $K$ and $K'$ points) that are induced by the loss difference between the two sublattice sets. The photonic lattice is constructed in a four-level $N$-type $^{85}$Rb atomic vapor cell with electromagnetically induced transparency (EIT) [30]. The instantaneous reconfigurability of such photonic lattice supports the demonstration of the wave behaviors under different parametric configurations. The honeycomb photonic lattice with the same complex refractive index for all lattice sites is "written" by a coupling field with a hexagonal intensity distribution under EIT in a three-level configuration [31-33]. The incident Gaussian probe beam is discretized into the honeycomb geometry by the induced lattice, and the light in two neighboring channels (corresponding to A and B sites) can experience power exchange due to evanescent coupling. With the introduction of another one-dimensional (1D) periodic pump field to cover one set of sublattice [for example, the A sublattice, see Figs. 1(a) and 1(c)] in the induced photonic structure, a four-level system is formed at the covered regions. Namely, the susceptibilities of A and B sublattices are resulted from the interactions between the probe beam and four- and three-level atomic configurations, respectively, leading to different transmission characteristics for the light inside corresponding waveguide channels. When the difference between the losses of A and B sites is large enough by changing the detuning of the 1D field, there emerge the local flat band surfaces which can localize the fields at one set of sublattice, suppressing the energy exchange between neighboring lattice sites. Such absorption-difference-induced flat band surface exhibits the similar zero-dispersion property as commonly seen in flat band structures such as Lieb lattices [27,28,34].

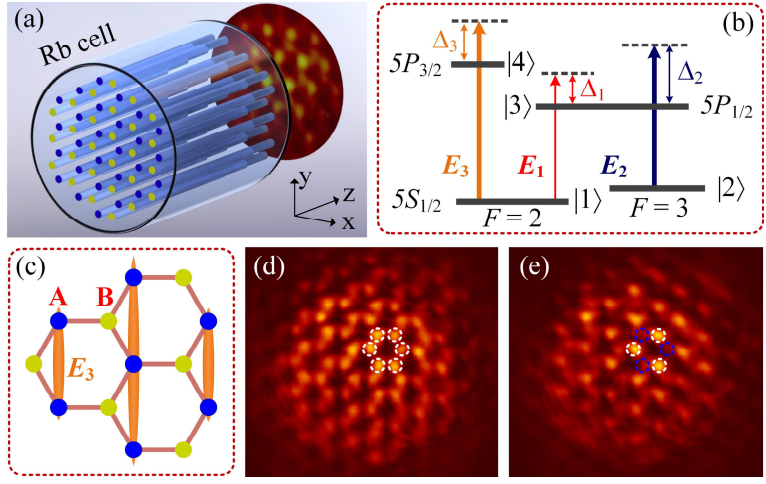

FIG. 1. (a) Sketch of the experimental principle. (b) The four-level $^{85}$Rb atomic configuration. (c) The spatial arrangements of the 1D periodic pump field $E_3$ and the induced honeycomb lattice inside the medium. (d-e) The observed output patterns of the probe field without and with A sublattice covered by the pump field, respectively.

The sketch of the experimental principle is shown in Fig. 1(a), where the honeycomb refractive index distribution inside the atomic vapor cell is optically induced by a coupling field $E_2$ with a hexagonal intensity profile (by interfering three coupling beams) under EIT. The lattice sites with different colors correspond to the A and B sublattices in the honeycomb photonic lattice. A Gaussian probe field $E_1$ is launched into the atomic vapor cell from the same side as $E_2$ to excite one $K$ valley, while the directions of the three coupling beams identify three $K'$ points in the reciprocal space. A 1D periodic pump field $E_3$ by interfering two pump beams is injected to cover the A sublattice. More details about the experimental settings are given in Supplemental Material [35].

As given in Fig. 1(b), the probe field $E_1$ together with the coupling field $E_2$ drives a three-level atomic configuration [$|1\rangle$ ($5S_{1/2}$, F=2)→$|3\rangle$ ($5P_{1/2}$)→$|2\rangle$ ($5S_{1/2}$, F=3)], in which an EIT window can be effectively created to enhance the refractive index felt by the probe field. The pump field $E_3$ drives the transition of $|1\rangle$→$|4\rangle$ ($5P_{3/2}$) to excite an $N$-type four-level atomic configuration (only at A sites), which can lead to different absorption for the two sets of sublattice. Detuning $\Delta_i$ ($i = 1, 2$ and 3) in the energy-level diagram represents the difference between the laser frequency and the frequency gap of the two levels connected by $E_i$. The spatial beam arrangement of the pump field and honeycomb photonic lattice is shown in Fig. 1(c). For the three-level EIT case without the 1D pump field, the resulted refractive index is inversely related to the coupling-field intensity under certain range of frequency detuning. With the two-photon detuning set as $\Delta_1-\Delta_2>0$, the reconfigurable non-Hermitian photonic lattice with honeycomb real and imaginary parts is effectively established [36]. The observed transmitted probe pattern (at the output surface of the cell and captured by a charge coupled device camera) under the condition of uniform susceptibility for both sublattices is shown in Fig. 1(d), which is clearly a discrete pattern in the honeycomb geometry. With the pump field turned on to modulate the imaginary part of A sublattice, the transmitted probe intensity [Fig. 1(e)] at A sites (marked by the blue dashed circles) is much weaker due to the larger imaginary part (representing stronger absorption).

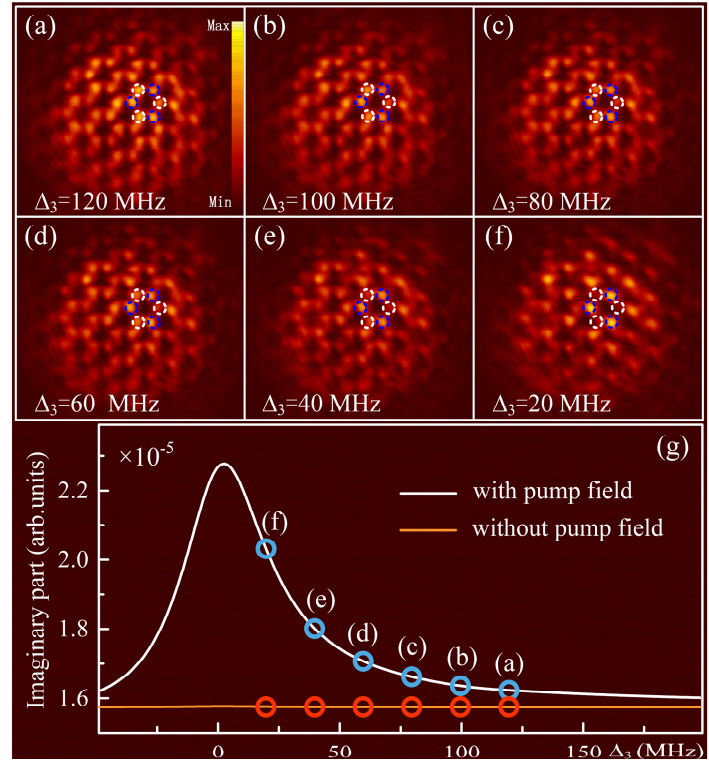

FIG. 2 (a)-(f) Detected field patterns by changing the pump detuning $\Delta_3$ from 120 MHz to 20 MHz. The two-photon detuning is fixed at $\Delta_1 - \Delta_2$=25 MHz; (g) The calculated imaginary parts of the susceptibility at A and B sites from the density matrix method. The six symbols correspond to the values of detuning in (a-f).

One advantage of the photonic lattices based on atomic coherence is the easily accessible tunability inherited from the multi-level atomic configuration. Here, we show that loss difference between A and B sites of the constructed non-Hermitian honeycomb lattice can be reconfigured by adjusting the laser parameters. Figures 2(a-f) demonstrate the evolution of the output intensity at A (blue circles) and B (white circles) sublattices versus the detuning of the pump field (which only affects the susceptibility of A sites) with other parameters fixed. With the pump detuning $\Delta_3$ discretely tuned from 120 MHz to 20 MHz, the experienced absorption in A sites becomes stronger, resulting in a decrease of the intensity inside blue circles, while the intensity at uncovered B sites governed by a three-level configuration exhibits no obvious change. These observed results clearly demonstrate the increase of the loss difference between A and B sublattices with decreasing $\Delta_3$.

The interaction between the probe beam and the four-level atomic configuration can be described by using the density matrix method with rotating-wave approximation [35, 37]. According to the density matrix equations, the loss parameters $\gamma_{a,b} = \text{Im}[\chi_{a,b}]$ at the A and B sites are plotted in Fig. 2(g), with symbols denoting the detunings corresponding to Fig. 2(a-f). As one can see, when $\Delta_3$ is tuned from 120 MHz to 20 MHz, $\gamma_b$ keeps constant while $\gamma_a$ increases, resulting in the enlargement of loss difference $\Delta\gamma = \gamma_a - \gamma_b$. The estimated loss difference values of Fig. 2(a-f) are

$\Delta\gamma = 0.49, 0.62, 0.88, 1.33, 2.27$, and $4.59$ ($\times 10^{-6}$), respectively. This result advocates the instantaneous reconfigurability of the established non-Hermitian honeycomb lattice, which in turn provides abundant space to govern the beam dynamics under various experimental conditions.

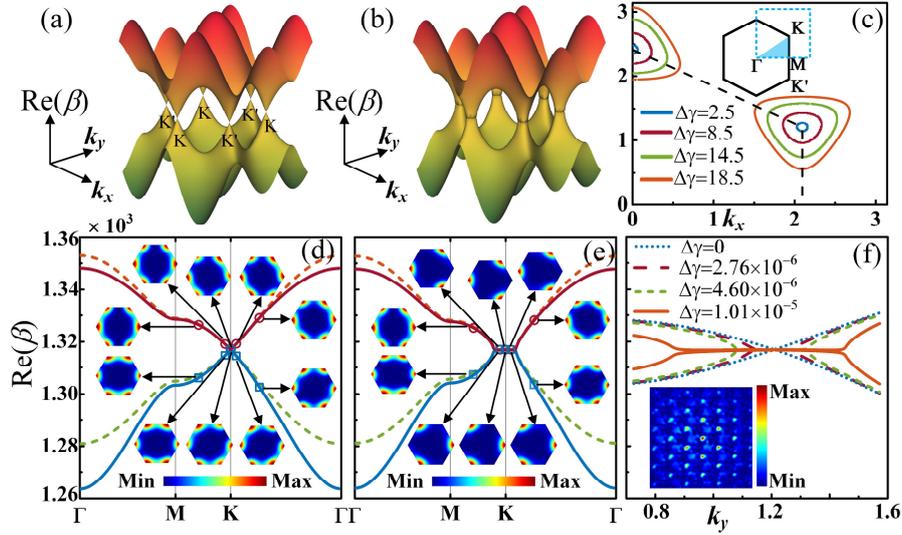

FIG. 3. Band structures and eigenstates of the non-Hermitian honeycomb lattice system. (a, b) Bands of Re[$\beta$] according to Eq. (2) for $\Delta\gamma = 0$ (a) and $\Delta\gamma = 8.5$ (b) with $\kappa = 12$. $n_0$ and $(\gamma_a + \gamma_b)/2$ can be set arbitrarily as they only shift the bands up or down. (c) Boundaries of the local flat band surfaces for different $\Delta\gamma$. The inset shows the first Brillouin zone. (d, e) Calculated band structures (solid lines) for the synthesized honeycomb lattice with $\Delta\gamma = 0$ (d) and $\Delta\gamma = 4.1 \times 10^{-6}$ (e), together with representative eigenstates (insets). The dashed lines are the corresponding bands in Eq. (2) from the analytic tight-binding model for comparison. (f) Evolutions of the local flat band around one $K$ point by increasing loss difference $\Delta\gamma$. The inset shows the localized field pattern when the local flat band (loss difference) is large enough for field localization.

Such a non-Hermitian honeycomb photonic lattice system can be described by the effective tight-binding model with Hamiltonian [38,39]:

$$\mathbf{H} = \begin{pmatrix} n_a & \kappa \sum_m e^{i\mathbf{k}\cdot\mathbf{r}_m} \\ \kappa \sum_m e^{-i\mathbf{k}\cdot\mathbf{r}_m} & n_b \end{pmatrix}, \quad (1)$$

where $n_{a,b} = n_0 - i\gamma_{a,b}$ are the normalized complex refractive indices at A and B sites, $\kappa$ is the normalized coupling strength between neighboring sites, $\mathbf{k} = (k_x, k_y)$ is the two-dimensional wavenumber, and $\mathbf{r}_{1,2,3} = (1,0), \left(-\frac{1}{2}, \frac{\sqrt{3}}{2}\right), \left(-\frac{1}{2}, -\frac{\sqrt{3}}{2}\right)$ are the normalized coupling directions on the $x-y$ plane. The eigenvalues of this Hamiltonian, which characterize the propagation properties of the probe beam inside the system, are found to be

$$\beta_{1,2} = n_0 - \frac{1}{2}i(\gamma_a + \gamma_b) \pm \kappa \sqrt{3 - \left(\frac{\Delta\gamma}{2\kappa}\right)^2 + 2\left(\cos\frac{3k_x}{2}\cos\frac{\sqrt{3}k_y}{2} + \cos(\sqrt{3}\,k_y)\right)}. \quad (2)$$

When there is no loss difference ($\Delta\gamma = \gamma_a - \gamma_b = 0$), such eigenvalues can be visualized in the band structure shown in Fig. 3(a). It is obvious that they degenerate at the Dirac points ($K$ and $K'$ points, real part of the eigenvalues) and locally show linear dispersion properties. However, when loss difference is introduced ($\Delta\gamma \neq 0$), the point degeneracy becomes circular-surface-like degeneracy and the original linear dispersion evolves into local dispersionless flat band around the K and K′ points. This can be seen in the band structure of Re[$\beta$] in Fig. 3(b) for $\Delta\gamma = 8.5$ and $\kappa = 12$ (normalized parameters). The corresponding bands of Im[$\beta$] are shown in Fig. S2(a) in Supplemental Material [35]. Such a local flat band surface can localize the incident light due to locally zero group velocity. Moreover, in Fig. 3(c), we plot the boundaries of the local flat bands for different $\Delta\gamma$. It is clear that with the increase of $\Delta\gamma$, the local flat band surfaces become broader, enabling better localization performance.

The formation of the local flat band under non-zero loss difference can be understood from the Dirac Hamiltonian approximation around the K and K′ points. By taking the Taylor series expansion of Eq. (1), one can obtain

$$\mathbf{H}_D = \begin{pmatrix} n_0 - i\gamma_a & \kappa' k e^{i\theta} \\ \kappa' k e^{-i\theta} & n_0 - i\gamma_b \end{pmatrix}, \quad (3)$$

where $\kappa' = 3\kappa/2$, $k$ is the amplitude of the wavenumber reference to K (K′) point, and $\theta$ is the local direction angle. The eigenvalues of this Dirac Hamiltonian are

$$\beta_{1,2} = n_0 - \frac{1}{2}i(\gamma_a + \gamma_b) \pm \kappa'\sqrt{k^2 - \left(\frac{\Delta\gamma}{2\kappa'}\right)^2}. \quad (4)$$

Now, the degeneracy condition $k^2 - (\Delta\gamma/2\kappa')^2 = 0$ is dependent on the amplitude of the local wavenumber $k$, indicating that the degenerate point evolves to an exceptional ring [23,40]. In addition, the size of the ring is linearly proportional to the loss difference $\Delta\gamma$. As the local flat band (real part of the eigenvalues) is inside the exceptional ring, the size of the local flat band increases proportionally to the magnitude of $\Delta\gamma$, as verified by Fig. 3(c).

We calculate the band structure of the synthesized photonic lattice with honeycomb susceptibility distributions using the PDE module of COMSOL Multiphysics and the results are shown as the solid lines in Fig. 3(d, e). One can see that the Dirac cone at K point for $\Delta\gamma = 0$ becomes local flat band surfaces for $\Delta\gamma = 4.6\times10^{-6}$, agreeing with the analytic results (dashed lines) of Eq. (2) from the tight-binding model, especially in the vicinity of K points. We note that the minor difference between bandstructures close to Γ point based on the two methods is due to the simplicity of the symmetric tight-binding model. To better demonstrate the formation of the local flat band surfaces, in Fig. 3(f), we plot the bands around K point for four different $\Delta\gamma$. With the increase of the $\Delta\gamma$ value, the local flat band expands broader. Also, we show representative eigenstates in the inset of Fig. 3(d, e). It is worth noting that, with the increase $\Delta\gamma$, the local flat bands not only emerge and expand, but also share the similar eigenstates for each band. Particularly, for the eigenstates of the upper (lower) local flat band, the field is localized at B (A) sites with low (high) loss. When a probe beam is obliquely incident into the system to excite the K vicinity, the eigenstates of both the local flat bands will be excited. However, since the B lattice sites have lower loss than the A sites, the eigenstates of field concentrated at A sites (lower band) will decay much faster and finally only the upper band survives, resulting in the field localization at B sites. The inset of Fig. 3(f) shows the simulated results of the localized field pattern when the probe beam propagates for a certain distance in the synthetic honeycomb photonic lattice with relatively large loss difference. The probe beam will keep this localized profile when propagating inside the lattice but experience the amplitude decay due to the loss.

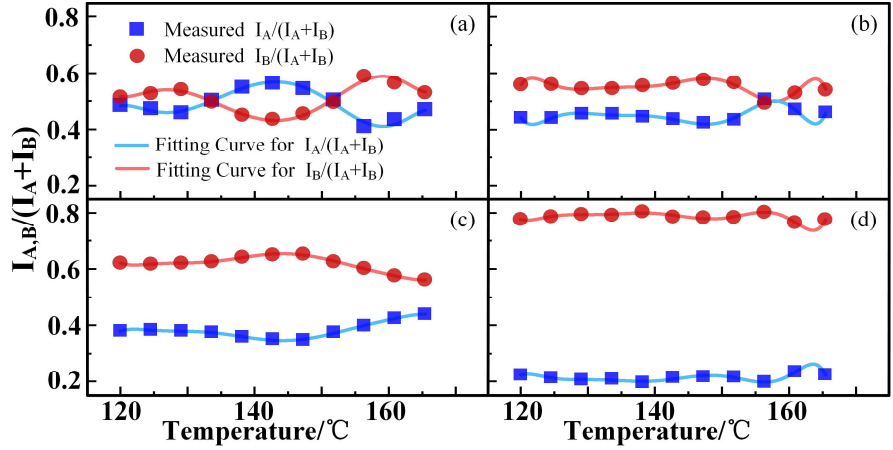

FIG. 4. Observed energy exchanges between sublattices A and B during the propagation of probe under different pump parameters: (a) without the 1D pump field $E_3$, (b) $\Delta_3 = 120$ MHz, (c) $\Delta_3 = 40$ MHz, and (d) $\Delta_3 = 20$ MHz. The squares and circles are the measured proportions of the transmitted intensities at A and B sites, respectively, while the corresponding solid curves represent mathematic fitting based on the experimental measurements and provide a guide to the eye.

In experiment, the field localization can then be verified by examining the energy exchange between A and B sites. Under the condition of a large enough loss difference, the field can be localized at B sites (three-level atomic region) and in principle there should be no energy transportation between A and B sites. While the field fails to be localized under the condition of a relatively small loss difference, the energy at A and B sites will couple to each other and resulting in energy fluctuation. In the following, we show the energy exchange between two adjacent waveguides (corresponding to A and B sublattices) at different $\Delta\gamma$ in the experiment. One advantage of atomic media is that increasing the atomic density (positively related to the temperature of the atomic vapor) is understood as extending the effective propagating distance of the probe beam inside the photonic lattice [31,41]. With the pump detuning adjusted to set different imaginary parts of the refractive index of the four-level atomic regions (A sites), the transport dynamics of the probe beam are captured in Fig. 4 when the medium is heated from 120 ℃ to 165 ℃. Actually, considering that the susceptibility is proportional to the atomic density $N$ for either the three- or four-level configuration [Eq. (S2) in the Supplemental Material [35], higher temperature can cause larger imaginary part as well as larger loss difference $\Delta\gamma$, which will result in weaker transmission but help to improve performance of the field localization. The relatively transmitted intensities $I_A$ and $I_B$ at A and B sites are measured by the software of the charge coupled device camera and their proportions are represented by symbols in Fig. 4.

When the pump field is absent, all the sites on the honeycomb lattice possess the same susceptibility controlled by a three-level configuration. In this case, the loss difference is zero ($\Delta\gamma = 0$) and the commonly seen power exchange between neighboring waveguide channels is observed in Fig. 4(a), where the initial intensities $I_A$ and $I_B$ are very close to each other. During the propagation, one can see clearly twice energy exchanges indicated by the two points where $I_A \approx I_B$.

With the small increase of $\Delta\gamma$ by setting the pump detuning at $\Delta_3 = 120$ MHz [see Fig. 2(g)], as shown in Fig. 4(b), the frequency of power exchange between A and B sites is reduced to only once at ~155°C, which means the energy exchange requires a longer propagating distance than the case without the pump field. The slowing down of the power exchange indicates the formation of the local flat band surface but still a limited area, which cannot fully localize the beam energy. With the pump detuning decreased to $\Delta_3 = 40$ MHz [Fig. 4(c)], the difference between the imaginary parts is further enlarged and a broader local flat band surface occurs. As a consequence, one can see that there is no power exchange in the given propagation distance. For the very large loss difference shown in Fig. 4(d) with $\Delta_3 = 20$ MHz, the ratio between the transmitted intensities $I_A$ and $I_B$ reaches ~1:4. Since the generated local flat band surface have a large enough area and can confine the energy mainly into B sublattices. Correspondingly, we simulate the beam propagation dynamics in the synthetic honeycomb photonic lattice with the susceptibility distribution at the same order of magnitude as the experiment [42]. The simulation results of the energy exchange depending on normalized propagating distance are plotted in Fig. S3, which shows very similar energy evolution behaviors as the experimental results.

In conclusion, we experimentally demonstrate the field localization behavior by switching the band structures from Dirac points with linear dispersion to dispersionless local flat band surfaces in a reconfigurable non-Hermitian honeycomb photonic lattice that induced in a multi-level atomic vapor cell. Such switching is derived from the induced loss difference between A and B sublattices, whose susceptibility are controlled by four- and three-level atomic configurations, respectively. By easily setting the laser parameters, the imaginary parts as well as the sizes of the formed local flat band surface are effectively manipulated to govern the beam dynamics in individual waveguide channel. The occurrence of the local flat band surface is verified by the vanishing of power exchange, indicating the light is localized by the flat-band modes. Being different from the phenomena relying on the degree of non-Hermiticity in previous works involving exceptional rings [19, 23-25], the localization of optical fields requires only the proper manipulation of loss difference. Namely, the localization behavior arising from loss difference is a result of the local flat band with shared eigenstates, and does not depend on the exceptional points. Our work not only uncovers a new property of non-Hermitian honeycomb photonic systems but also opens the door for the experimental exploration on the capabilities of rings-surrounded flat bands in high-dimensional non-Hermitian systems, promisingly extending to quantum, mechanical, and electrical non-Hermitian systems.

This work was supported by National Key R&D Program of China (2018YFA0307500), National Natural Science Foundation of China (62022066, 12074306, 11804267), and the Key Scientific and Technological Innovation Team of Shaanxi Province (2021TD-56).

[*]Y. F. and Z. L. contributed equally to this work.

[†]fu.liu@xjtu.edu.cn

[‡]zhyzhang@xjtu.edu.cn